# Optimal Taxation of Assets

Nicolaus Tideman and Thomas Mecherikunnel

Abstract

The optimal taxation of assets requires attention to two concerns: 1) the elasticity of the supply of assets and 2) the impact of taxing assets on distributional objectives. The most efficient way to attend to these two concerns is to tax assets of different types separately, rather than having one tax on all assets. When assets are created by specialized effort rather than by saving, as with innovations, discoveries of mineral deposits and development of unregulated natural monopolies, it is interesting to consider a regime in which the government awards a prize for the creation of the asset and then collects the remaining value of the asset in taxes. Analytically, the prize is like a wage after taxes. In this perspective, prizes are awarded based on a variation on optimal taxation theory, while assets of different types are taxed in divergent ways, depending on their characteristics. Some categories of assets are abolished.

## I. Introduction

When economists develop a framework for the optimal taxation of assets, they often assume that all assets are aggregated into a single category and that taxation will be proportional to the income that the assets generate.[1] We depart from both assumptions, assuming instead that assets in different categories will be taxed according to separate rules, and that taxes will resemble property taxes rather than income taxes, being proportional to the value of assets rather than to the income that assets generate.

We divide assets into six categories:

1) Land and other useful privileges
2) Useless privileges
3) Capital
4) Intellectual property
5) Mineral deposits

[1] Mankiw, N. Gregory, Matthew Weinzierl, and Danny Yagan. "Optimal taxation in theory and practice." *Journal of Economic Perspectives* 23, no. 4 (2009): 147-74.

6) Unregulated natural monopolies

We could consider a seventh category of regulated natural monopolies, but taxing them is relatively uninteresting, because it is generally more efficient to restrict the income of the owners of regulated monopolies by regulating the prices they charge than by taxing their income or the value of their assets. If it is efficient to tax the income or the value of the assets of the owners of regulated monopolies, then one must consider the interaction of regulation and taxation, which introduces a level of complexity that is not present in the categories of assets that we consider.

Taxation of assets according to the income that they generate usually has a simplicity that is not present when assets are taxed according to their value. Income is an observable flow, while the value of an asset is not so observable. Still, income is not as transparently observable as one might suppose. Depreciation must be subtracted to calculate economic income, and that is generally not observable. Furthermore, if one wishes to tax asset income according to a different schedule than labor income, there are difficulties in applying the rules to self-employed persons who combine their labor with assets that they own. And setting a value of assets for tax purposes is possible. Property tax assessors have been doing it for centuries.

Taxing assets according to their value rather than their income has the virtue of tilting taxes away from those who make efficient use of their assets and toward those who make poor use of their assets, providing greater encouragement than under an income tax for the transfer of assets to those who can make good use of them. Thus, we believe it is worthwhile to specify a tax on assets as proportional to their value rather than proportional to the income from them.

In the sections that follow, we first consider the categories of assets one at a time and then consider the form of a tax that is optimal for all of them.

## II. Land and other useful privileges

The word "privilege" comes from the Latin words, *prive* (private) and *lege* (law). A private law is a law that has someone's name in it, a law permitting one person to do what others are not permitted to do. That is, a private law creates a privilege. Land titles are privileges because they assign, to particular private parties, exclusive rights to opportunities that have value that is not the result of private effort. The rent that can be collected from a useful privilege could go to anyone, without reducing the incentive to engage in productive activity.

Land titles are *useful* privileges when land yields the greatest value when *someone* has exclusive rights of use, rather than everyone having unrestricted use, as with removing salt from the oceans. Other examples of useful privileges are rights to exclusive use of portions of the radio frequency spectrum and rights to take-off and landing slots at airports. Useful privileges represent scarce opportunities that, though they are created without effort by those who hold them, need to be allocated by the market for production to be efficient.

Since the time of the Physiocrats and Adam Smith, economists have understood that a properly administered tax on land or other privileges has no harmful effect on incentives to be productive. As Smith put it,

> Both ground-rents and the ordinary rent of land are a species of revenue which the owner, in many cases, enjoys without any care or attention of his own. Though a part of this revenue should be taken from him in order to defray the expenses of the state, no discouragement will thereby be given to any sort of industry. The annual produce of the land and labour of the society, the real wealth and revenue of the great body of the people, might be the same after such a tax as before. Ground-rents and the ordinary rent of land are, therefore, perhaps, the species of revenue which can best bear to have a peculiar tax imposed upon them.[2]

In modern terminology, a properly administered tax on land or other useful privileges is a "lump-sum" tax, that is, a tax with a magnitude that does not vary with any action taken by the owner of the land or other privilege. A lump-sum tax on an asset has no excess burden if the magnitude of the tax to be collected is independent of any action that the taxpayer might take, and the tax is less than the value of using the asset. Another way that this point is sometimes put is that the elasticity of the supply of land, properly defined, is zero, so that, unlike the case when an elasticity of supply is positive, a tax on land has no effect on the quantity of land that is available and therefore has no excess burden.

It might seem that the possibility of selling an asset makes a tax on holding an asset not a lump-sum tax. However, if the tax goes with the asset, the magnitude of the tax is not affected by its sale, so the condition of independence of action by taxpayer is still satisfied.

Since a tax on land or other useful privileges is a lump-sum tax, all that prevents the tax from taking all of the rental value of the asset, without any excess burden, is the imperfection of assessment. If the tax bill exceeds the rental value of an asset, then the owner of that asset will have an incentive to abandon the asset and disappear, if that is possible. As long as there is no danger of the tax bill exceeding the rental value of an asset, and the tax is independent of any action by the owner of the asset, there is no excess burden from the tax, no matter what fraction of the income from the asset, less than 100%, the tax takes.

---

[2] Smith, Adam. *The Wealth of Nations*, Book V, Chapter 2, pp. 795-96 in the Modern Library edition.

There is some mathematics associated with a property tax that is worth mentioning. The sale value of an asset is the present value of the net income from the asset after taxes, where income includes the consumption value of an asset that is held for consumption purposes. Thus, for an asset with a constant flow of pre-tax income, $y$, a constant flow of taxes, $T$, and the current time denoted by 0, the value is given by

$$V = \int_0^\infty (y - T)e^{-\rho\tau} d\tau = \frac{y-T}{\rho}, \tag{1}$$

where $\rho$ is the discount rate for flows of income after taxes and $\tau$ is time. If the tax is defined as the percentage $t$ of the value of the asset, then $V = \frac{y-tV}{\rho}$, implying that $V = \frac{y}{t+\rho}$. In words, the value of an asset with a constant stream of income and a tax proportional to its value is the pre-tax income of the asset, discounted at the sum of the tax rate and the discount rate for after-tax flows of income.

It might seem that there is an upper limit of 100% on the tax rate, but this is not true, provided that tax payments can be sufficiently frequent. A tax rate of 2% per year is a rate of 200% per century. The formula for value assumes implicitly that tax payments are continuous. Thus it would be possible to have a tax rate of 15% per month, which would be 180% per year. If the interest rate is 6% per year, or 0.5% per month, then a tax rate of 15% per month of the value of a privilege would take $\frac{\frac{V}{0.5} - \frac{V}{15.5}}{\frac{V}{0.5}}$ or 30/31 of the value of the privilege. For any share of value less than 100%, there are combinations of tax rates on value and frequencies of payment that would take that share of value.

One apparently limiting factor on the share of value that a tax on the value of a privilege can take is the institution of mortgages. If a tax causes the value of an asset to fall below the equity that the owner of an asset has in the asset after the mortgage on it, then the owner has an incentive to abandon the asset, which would create an excess burden. To avoid this limiting factor, part of the tax bill can be sent to the owner of the mortgage.

The coherence of doing this comes from the fact that by recording a mortgage at a courthouse and preventing an asset from being sold without satisfying the mortgage, a mortgagor is declaring himself to be a partial owner of that asset. Therefore it is reasonable, when a taxing jurisdiction decides to levy taxes on privileges, to treat a mortgagor as an owner to be taxed, for the share of the asset that he owns. Thus, the institution of mortgages need not prevent a jurisdiction from taking, through a tax proportional to the value of privileges, an arbitrarily large share of the value of those privileges.

**III. Useless privileges**

Some privileges subtract from rather than add to a nation's productivity. Examples would be acreage allotments, import quotas and taxi medallions. These privileges rewarded their original owners for successful rent-seeking. No excess burden is imposed on an economy by reducing the value of these privileges by taxation. As with useful privileges, a tax proportional to their value can take an arbitrarily large share of the income they create. When no efficiency purpose is served by restricting the quantity of these activities, it is most efficient to simply abolish these privileges. When there is an efficiency purpose to restricting quantity, efficient management is achieved by setting a price equal to marginal social cost and allowing anyone who wishes to do so to engage in the restricted activity.

**IV. Capital**

The meaning of the term "capital" has evolved over the centuries. In the 19th century, capital described a class of assets distinct from land. Thus, David Ricardo defined capital as "that part of the wealth of a country which is employed in production, and consists of food, clothing, tools, raw materials, machinery, etc., necessary to give effect to labor."[3] John Stuart Mill wrote, "Whatever things are destined to supply productive labor with the shelter, protection, tools and materials which the work requires, and to feed and otherwise maintain the laborer during the process, are capital."[4] Henry George defined capital as "wealth in the course of exchange,"[5] and defined wealth as "natural products that have been secured, moved, combined separated, or in other ways modified by human exertion, so as to fit them for the gratification of human desires,"[6] thus excluding land from the definition of wealth and hence from the definition of capital.

---

[3] Ricardo, David. Principles of Political Economy, Chapter V.
[4] Mill, John Stuart. Principles of Political Economy, Book I, Chap. IV.
[5] George, Henry. Progress and Poverty, Book I, Chapter 2, page 48 in many editions.
[6] Ibid., pp. 41-42.

Around the turn of the 20th century, economists began to include all assets in the definition of capital,[7] thereby eliminating the possibility of a simple name for the category described by the 19th meaning of "capital." Since our analysis requires precisely that category, we return to the 19th century definition and define capital as valuable physical things that are produced by combining natural opportunities, labor, and previously produced physical things.

The traditional reasoning with respect to taxation of capital is that it is most efficient to not tax it at all.[8] However, Saez and Stantcheva have pointed out that this conclusion rests on an assumption of uniformity among taxpayers with respect to the ratio of labor income to asset income, and that when this uniformity does not obtain and the social welfare function gives different weights to persons with different ratios of labor income to capital income, social welfare can be increased by some taxation of the aggregate of all assets including capital.[9] While this is true, the capacity of a recurring tax on capital to add to social welfare is limited by positive elasticity of the supply of capital, that is, by the ability of the owners of capital to decline to renew it as it wears out. The greater the elasticity of the supply of capital, the smaller is the opportunity to increase social welfare by taxing capital. As long as there are incompletely used sources of public revenue that do not entail any excess burden, like the taxation of land and other useful privileges, it is likely that social welfare would be increased by replacing any existing recurring tax on capital with a tax on useful privileges that yielded the same revenue.[10] This would fail to be true only if the social welfare weights of the owners of useful privileges were so much greater than the social welfare weights of the owners of capital that the efficiency disadvantages of taxing capital were outweighed.

It is possible to tax capital without any excess burden if one can subject capital to a one-time unanticipated levy that does not induce an expectation with a positive probability of any future levy. However, it is notably difficult to meet this condition. If it would be possible to meet the condition, it would probably be by something like instituting reparations for the history of slavery, where there is a moral justification for the levy that does not appear to have a future counterpart.

---

[7] For a discussion of this evolution in the definition of capital and an argument that it was intended to obscure the ideas of Henry George, see Gaffney, Mason. "Neo-classical economics as a stratagem against Henry George," in Gaffney, Mason, and Fred Harrison. *The Corruption of Economics*. London: Shepheard-Walwyn, 1994: 29-163.

[8] Atkinson, Anthony Barnes, and Joseph E. Stiglitz. "The design of tax structure: direct versus indirect taxation." *Journal of Public Economics* 6, no. 1-2 (1976): 55-75.

[9] Saez, Emmanuel, and Stefanie Stantcheva. "A simpler theory of optimal capital taxation." *Journal of Public Economics* 162 (2018): 120-142.

[10] For empirical evidence on this point, see Tideman, T. Nicolaus and Florenz Plassmann, "A Markov Chain Monte Carlo Analysis of the Effect of Two-Rate Property Taxes on Construction," *Journal of Urban Economics* 47 (March 2000) pp. 216-47.

**V. Intellectual property**

Intellectual property generates inefficiency from the monopoly power that it provides. It would be more efficient to motivate the creation of intellectual property with prizes than with monopolies, if it is possible to estimate the value of innovations with adequate accuracy and to obtain funding for the prizes with adequate efficiency. For the purpose of analysis, we make these assumptions. We assume that taxation of land and other useful privileges will provide enough revenue for efficient prizes for innovation. Then the specification of the optimal prize structure is analogous to optimal wage taxation.

The value generated by an innovation is like the income from effort. The portion of that value not going to the innovator is like tax revenue, except that it goes to the public instead of going into the public treasury. The prize is equivalent to the share of wage income that a worker has after taxes.

Some innovations occur without significant cost; others involve huge costs. It seems sensible to have a prize structure that starts by awarding innovators all of the value generated by their innovations, up to some multiple of their costs, including interest. If one was prepared to compensate all unsuccessful would-be innovators for all of their costs, then that multiple might reasonably be 1.0. However, it seems likely to us that it would not be attractive to compensate all unsuccessful would-be innovators for all of their costs, so we assume that to compensate for the probability of unsuccessful attempted innovations, the prize for an innovation would start with something like the smaller of the value of the innovation and three times its cost. If the value was more than three times the cost, then one would apply a theory of optimal prizes that is isomorphic with the theory of optimal taxation of wage income.

For this purpose, we adapt the theory of optimal taxation of wages and assets in Saez and Stantcheva (2018). In their framework, at any time, the rate of utility accumulation of person $i$, $u_i$, is $u_i\,(c, k, z) = c + a_i(k) - h_i(z)$, where $c$ is consumption, $a_i(k)$ is an increasing, concave function of wealth, $k$, and $h_i(z)$ is the (increasing, convex) disutility of earning labor income, $z$.

To keep things simple, we assume a disjunction between persons who earn wages and persons who earn prizes, so that we can use equations similar to those used by Saez and Stantcheva to specify the optimal prize. In particular, define $h_{2i}(s)$ as the disutility to person $i$ of generating an innovation with a flow of value of $s$ per unit of time. Then when person $i$ is someone who earns prizes rather than wages, her utility function becomes

$$u_i(c, k, z) = c + a_i(k) - h_{2i}(s). \tag{2}$$

The present value of utility from the combination of the paths of consumption, wealth possession, and production of innovations $\{c_i(t), k_i(t), s_i(t)\}_{t\geq 0}$ is:

$$V_i(\{c_i(t), k_i(t), s_i(t)\}_{t\geq 0}) = \delta_i \int_0^{\infty} [c_i(t) + a_i(k_i(t)) - h_{2i}(s_i(t))] e^{-\delta_i t} dt. \tag{3}$$

The normalization by $\delta_i$ implies that in additional unit of consumption in perpetuity increases utility by one unit for all persons. Prior to the imposition of taxes, person $i$ has wealth $k_i^{init}$.

With no taxes on capital, the instantaneous budget constraint of person $i$ is:

$$\frac{dk_i(t)}{dt} = n_i + rk_i(t) + P_N(s_i(t)) - c_i(t). \tag{4}$$

where $n_i$ is person $i$'s income after taxes from land and other useful privileges, $r$ is the rate of return on capital, and $P_N(s)$ is the prize per unit of time for an innovation with a rate of value generation of $s$ per unit of time.

The Hamiltonian of individual $i$ at time $t$, with co-state $\lambda_i(t)$ on the budget constraint, is:

$$H_i(c_i(t), s_i(t), k_i(t), \lambda_i(t) = c_i(t) + a_i(k_i(t)) - h_{2i}(s_i(t))$$
$$+\lambda_i(t) \cdot [n_i + rk_i(t) + P_N(s_i(t)) - c_i(t)]. \tag{5}$$

The first-order conditions imply that the choice of $(c_i(t), k_i(t), s_i(t))$ must be such that

$$\lambda_i(t) = 1,$$
$$h_{2i}'(s_i(t)) = P_N'(s_i(t)),$$
$$a_i'(k_i(t)) = \delta_i - r, \quad \text{and}$$
$$c_i(t) = n_i + rk_i(t) + P_N(s_i(t)). \tag{6}$$

Because utility is defined to be proportional to consumption, the variables wealth, and income jump immediately to their steady-state values, characterized by

$$h_{2i}'(s_i(t)) = P_N'(s_i(t)),$$
$$a_i(k_i(t)) = \delta_i - r, \quad \text{and}$$
$$c_i(t) = n_i + rk_i(t) + P_N(s_i(t)). \tag{7}$$

Therefore

$$V_i(\{c_i(t), k_i(t), s_i(t)\}_{t\geq 0}) = [c_i(t) + a_i(k_i(t)) - h_{2i}(s_i(t))] + \delta_i(k_i - k_i^{init}) \quad (8)$$

The last term represents the benefit (or cost if negative) of the introduction of redistributive taxes and transfers.

The derivative of the optimal non-linear tax on wages is specified (Saez and Stantcheva, 2018, p. 123) by

$$T_L'(z) = \frac{1-\bar{G}_L(z)}{1-\bar{G}_L(z)+\alpha_L \cdot e_L(z)}, \quad (9)$$

where:

$\bar{G}_L(z)$ is the average relative welfare weight for persons with labor income greater than $z$.

$\alpha_L(z) = \frac{z \cdot h_z(z)}{1-H_z(z)}$. is the local Pareto parameter of the labor income distribution, where:

$H_z(z)$ is the cumulative distribution of labor income, and

$h_z(z)$ is the corresponding density when the tax system is linearized at $z$.[11]

Finally, $e_L(z)$ is the local elasticity of the supply of labor with respect to $1 - T'$ at $z$.

To adapt this tax structure to prizes, we apply the fact that when an innovator receives a prize rather than a monopoly, the prize can be conceived as what is left after taxes, while the tax is like the value of the innovation minus the prize, except that the value of the innovation is distributed among the public rather than going into the public treasury. Thus, the derivative of the optimal non-linear prize for an innovation, with respect to the value of the innovation, is

$$P_N{}'(s) = \frac{\alpha_N(s) \cdot e_N(s) G^*}{1-\bar{G}_N(s)+\alpha_N(s) \cdot e_N(s) G^*}, \quad (10)$$

where:

$\bar{G}_N(s)$ is the average relative welfare weight for persons who produce innovations worth more than $s$,

$\alpha_N(s) = \frac{s \cdot j_S(s)}{1-J_S(s)}$ is the local Pareto parameter of the distribution of the value of innovation production across individuals, where:

$J_S(s)$ is the cumulative distribution of innovation production, and

$j_S(s)$ is the corresponding density when the prize system is linearized at $s$,

[11] Here Saez and Stantcheva give the letter *h* a second meaning, unrelated to the first meaning that they give it. We use the letter *j* for this second meaning, to avoid confusion.

$e_N(s)$ is the local elasticity of the supply of innovations with respect to prizes for innovations, at $s$,

And finally, $G^*$ is the welfare weight of the recipient of the average dollar of benefits from innovations, relative to the average person in the society.

Since the value of an innovation is often not obvious when the innovation is first announced, it would be sensible to have prizes that reflected estimates of benefits on a period-by-period basis.

The two important effects of rewarding innovations with prizes rather than monopolies are that everyone would have free, unfettered access to all innovations and that innovators would be rewarded optimally, taking account of a) the elasticity of innovations with respect to prizes and b) the relative welfare weight for innovators, relative to those who benefit from innovations.

### VI. Mineral Deposits

Mineral deposits have in common with innovations the fact that their discovery requires effort from persons with special abilities. As with innovations, more discoveries of mineral deposits can be expected if those who discover them receive greater rewards. At the same time, there is value in economizing on rewards, since rewards subtract from public revenue.

Unlike innovations, mineral deposits are generally most valuable when access to them is restricted to a single party. In this they are like land and other useful privileges. Thus, mineral deposits provide a possible source of public revenue, from assigning rights to them to the highest bidders. (While the United States assigns ownership of mineral deposits to those who own surface rights, many other countries assign ownership of mineral rights to the government. Such assignment of mineral rights to the government is needed to make it worthwhile to award prizes for discoveries.)

The discovery of a mineral deposit has a general equilibrium effect, potentially changing all of the prices in an economy. We treat this general equilibrium effect as distributionally neutral in terms of the social welfare function, so that the benefit of a mineral discovery, in terms of the social welfare function, is the public revenue generated by taxing access to the deposit, plus any prize or other profit going to the discoverer, adjusted for the relative social welfare weight of the discoverer.

The structure of the optimal prize for the discovery of a mineral deposit is much like that of the optimal prize for an innovation, with the difference that, because the benefit, apart from the prize, takes the form of public revenue rather than public benefits, there is no adjustment for the average social welfare weight of the recipients of benefits. Thus, assuming that those who receive prizes for the discovery of mineral deposits receive neither wages nor prizes for innovations, and that the whole value of the mineral deposit, after the payment of the prize is collected by the government, the derivative of the optimal non-linear prize for discovery of mineral deposits, with respect to the value of the mineral deposit, is:

$$P_M{}'(u) = \frac{\alpha_M(u) \cdot e_M(u)}{1 - \bar{G}_M(u) + \alpha_M(u) \cdot e_M(u)}, \tag{11}$$

where:

$\bar{G}_M(u)$ is the average relative welfare weight for persons who discover mineral deposits worth more than $u$,

$\alpha_M(u) = \frac{u \cdot j_u(u)}{1 - J_u(u)}$ is the local Pareto parameter of the distribution of the value of mineral deposit discoveries across individuals, where:

$J_u(u)$ is the cumulative distribution of the value of mineral deposit discovery, and

$j_u(u)$ is the corresponding density when the prize system is linearized at $u$,

And $e_M(u)$ is the local elasticity of the supply of mineral deposit discoveries with respect to prizes for mineral deposit discoveries, at $u$.

As with innovations, one would probably want to start the prize for discoveries of mineral deposits with something like the smaller of the value of the deposit and three times the cost of discovery, to compensate for all of the unsuccessful effort to discover mineral deposits.

The prize described by equation (11) could be paid as a lump sum upon discovery if the value of a discovery was readily apparent and the deposit would be exhausted relatively quickly. If the value of a mineral deposit could learned only as the deposit was worked, or if the mineral would be extracted over many years, then one would want to implement recurring prize payments and recurring payments to the government for access and for depletion. The combination of prize payments and payments for access and depletion would maximize the contribution of mineral deposits to social welfare.

**VII. Unregulated natural monopolies**

The economic theory of competition concludes that a competitive business will be worth the present value of net investment in the business, defined as the present value of investment less the present value of dividends. If a business is worth more than the present value of net investment, then the excess value is evidence of an absence of perfect competition in the activity in which the business operates. On this basis, we call a business that is worth more than the present value of net investment an "unregulated natural monopoly." Unregulated natural monopolies are like mineral deposits, in that special talent is required to find them. If ordinary effort could replicate them, they would yield only ordinary returns. Amazon and Facebook are prominent examples of unregulated natural monopolies.

The fact that an unregulated natural monopoly yields a persistent return greater than the ordinary return to investment provides an opportunity to raise public revenue by a recurring charge on the ownership of this type of asset, while optimally motivating discovery of such opportunities by prizes.

The structure of the possibility of identifying a previously unnoticed opportunity for an unregulated natural monopoly is much like the structure of the possibility of finding a previously unnoticed mineral deposit, except for the nature of what is hidden. Where an aspect of nature is hidden in the case of a mineral deposit, an aspect of potential social organization is hidden in the case of an unregulated natural monopoly. Because the structures are much the same, the optimal taxes and optimal prizes are much the same. One important difference is that mineral deposits are intended to be depleted, while unregulated natural monopolies are generally not. This implies that the optimal taxation of mineral deposits entails severance charges, while there is no corresponding feature of the optimal taxation of unregulated natural monopolies.

Assuming that those who receive prizes for the development of unregulated natural monopolies receive neither wages nor prizes for innovations nor prizes for discoveries of mineral deposits, and assuming that the full value of the unregulated monopoly after the payment of the prize is collected by the government, the derivative of the optimal non-linear prize for the development of an unregulated natural monopoly, with respect to the value of the unregulated natural monopoly, is:

$$P_R{}'(v) = \frac{\alpha_R(v)\cdot e_R(v)}{1-\bar{G}_R(v)+\alpha_R(v)\cdot e_R(v)}, \tag{10}$$

where:

$\bar{G}_R(v)$ is the average relative welfare weight for persons who develop unregulated natural monopolies worth more than $v$,

$\alpha_R(v) = \frac{v\cdot j_v(v)}{1-J_v(v)}$ is the local Pareto parameter of the distribution of the value of unregulated natural monopolies across individuals, where:

$J_v(v)$ is the cumulative distribution of the value of unregulated natural monopolies,

$j_v(v)$ is the corresponding density when the prize system is linearized at $v$,

And $e_R(v)$ is the local elasticity of the supply of unregulated natural monopolies with respect to prizes for their development, at $v$.

The idea of taxing the excess value of unregulated natural monopolies has a precedent in the idea that profits can be taxed without incurring an excess burden.[12] That literature, however, generally assumed that profit opportunities arise without effort and with essentially a zero elasticity of supply. The prize proposed for development of unregulated natural monopolies takes account of the positivity of the elasticity of supply.

The tax under consideration in this paper is not a profits tax, but rather a recurring "property" tax on the excess value of a business, that is, a recurring tax on the amount by which the value of a business exceeds the present value of net investment in the business.

**VIII. Valuation**

There are several issues of valuation that arise with respect to the tax system we analyze. First, there is the issue of measuring the costs of the things for which prizes are provided. Because of the difficulty of assigning an opportunity cost to the time of an independent prospector or an author, as well as the difficulty of identifying the amount of time spent in a creative effort, some conventions are likely to be needed. These might be such conventions as: The time of any prospector is worth \$500 per day. The cost of a written work is \$1 per word for general prose, \$2 per word for children's prose, and \$5 per word for poetry. Music costs \$1 per beat. Two-dimensional visual art costs \$1 per square inch. These are the costs that would be multiplied by some factor like three (to compensate for unsuccessful efforts) and then fully reimbursed if the activity for which a prize was given created that much value.

Then there is the issue of assigning value to the things for which prizes would be awarded. For music and writings, downloads and printed copies could be counted. For drugs, the number of patients treated could be counted. For inventions, the number of copies sold could be counted. These numbers would need to be combined with analysis that estimated the average consumer surplus per use, with further rewards for further value.

---

[12] Diamond, Peter A., and James A. Mirrlees. "Optimal taxation and public production I: Production efficiency." *The American Economic Review* 61, no. 1 (1971): 8-27.

For unregulated natural monopolies that were not publicly traded, there could be a system of competitive assessment, with an award of, say, 1% of the taxes collected, for the person who made the highest bid for the company. For publicly traded companies, the unregulated natural monopoly element in the value of the company would be specified as the greater of the market value of the company and the highest take-over bid, minus the present value of net investment.

**IX. Proposed Taxes**

We propose that there be no taxes on useless privilege or intellectual property, because they would simply be abolished. There would be no tax on capital, except possibly a one-time lump-sum tax supported by an ethical argument for redistribution. The value of mineral deposits above prize payments would be collected publicly by the combination of auctions and severance fees. Land would be taxed by a variation on the current property tax, with assessed values determined by something like current assessment methods and tax rate that would be quite high by historical standards. The tax rate might be something like 5% per month. In conjunction with an interest rate of 6% per year (0.5% per month) a tax of 5% of value per month would collect 10/11 of the rent of land publicly. For unregulated natural monopolies, the base of the tax would be the market value of the unregulated natural monopoly minus the present value of net investment.

Table 1: Optimal Taxes on Assets by Type of Asset

| | |
|---|---|
| Land and other useful privileges | Collect up to nearly all of the rental value of land and other useful privileges. |
| Useless privileges | Abolish useless privileges. |
| Capital | Leave capital untaxed, unless there is a basis in justice for a one-time capital levy. |
| Intellectual property | Replace intellectual property with optimal prizes. |
| Mineral deposits | Reward discovery of mineral deposits with optimal prizes, then collect up to nearly all of the remaining value with a combination of property taxes and severance taxes. |
| Unregulated natural monopolies | Reward discovery of unregulated natural monopolies with optimal prizes, then collect up to nearly all of their excess value with a property tax. |

The tax rate would be the same as the tax rate on land, collecting the same share of the value of the natural monopoly. These recommendations are summarized in Table 1.

## X. Conclusion

The different treatments of different categories of assets are related to the characteristics of assets, as shown in Table 2. Useless privilege and intellectual property are efficiently eliminated because restricted access is not efficient for them. Capital is left untaxed because its value comes entirely from human effort plus natural opportunities on which appropriate taxes would have been paid. The three categories that are taxed, land and other useful privileges, mineral deposits, and unregulated natural monopolies, share the characteristics that restricted access in efficient and there is a source of value other than human effort. The three assets for which prizes are awarded, intellectual property, mineral deposits, and unregulated natural monopolies, share the characteristic that their value comes, in part, from entrepreneurship.

Table 2: Characteristics of Assets that Affect their Optimal Taxation

| | Sources of value | | | Is restricted access efficient? |
|---|---|---|---|---|
| | Nature or Society (Rent) | Human Effort (Wages) | Discovery (Entrepre-neurship) | |
| Land and other useful privileges | Yes | | | Yes |
| Useless privileges | Yes | | | |
| Capital | Yes | Yes | | Yes |
| Intellectual property | | Yes | Yes | |
| Mineral deposits | Yes | Yes | Yes | Yes |
| Unregulated natural monopolies | Yes | Yes | Yes | Yes |